\documentclass[runningheads]{llncs}

\usepackage[utf8]{inputenc}
\usepackage{graphicx}
\usepackage{xcolor}
\usepackage{amsmath}
\usepackage{amsfonts}
\usepackage{url}
\usepackage{caption}
\usepackage{subcaption}
\usepackage{xcolor}

\title{A Ptolemaic Partitioning Mechanism}

\author{Richard Connor}

\institute{University of St Andrews, St Andrews, Scotland\\
\email{rchc@st-andrews.ac.uk}}

\begin{document}

\maketitle
\begin{abstract}

For many years, exact metric search relied  upon the property of triangle inequality to give a lower bound on uncalculated distances. Two exclusion mechanisms  derive from this property, generally known as pivot exclusion and hyperplane exclusion. These mechanisms work in any proper metric space and are the basis of many metric indexing mechanisms.
More recently, the Ptolemaic  and four-point lower bound properties have been shown to give  tighter  bounds in some subclasses of metric space.

Both triangle inequality and the four-point lower bound  directly imply straightforward \emph{partitioning} mechanisms: that is, a  method of dividing a finite space according to a fixed partition, in order that one or more classes of the partition can be  eliminated from a search at query time. However, up to now, no  partitioning principle has   been identified for the Ptolemaic inequality, which has    been used only as a filtering mechanism.

Here,  a novel partitioning mechanism for the Ptolemaic lower bound is presented. It is always better than either pivot or hyperplane partitioning. While the exclusion condition itself is weaker than Hilbert (four-point) exclusion, its calculation is cheaper. Furthermore, it can be combined with Hilbert exclusion  to give  a new  maximum for exclusion power with respect to the number of distances measured per query.

\end{abstract}
\keywords{Metric Search \and Partitioning  \and Ptolemaic inequality \and Supermetric space}
\section{Background and Related Work}

This article concerns querying a large finite space $(S,d)$ which is a subset of an infinite metric space $(U,d)$.

In most general terms, querying the space $(S,d)$ with query $q \in U$ is the task of finding a subset $\{s \leftarrow S \,|\, d(q,s) \le t\}$, for some value $t$ which gives a suitable size of solution set.
\begin{table}[]
\caption{Notation used throughout}
\begin{center}
\begin{tabular}{|l|l|}
\hline
Symbols&Meaning\\
\hline
$(U,d)$	&	An infinite metric space with domain $U$ and distance $d$\\
$(S,d)$		&	A large finite  space $S \subset U$ over which search is performed\\
$P$			& 	A small reference set $P \subset U$, usually $P \subset S$ \\
$u, u_0, u_1,\dots$		&	Elements of the infinite domain $U$\\
$s, s_0, s_1, \dots$		&	Elements of the finite domain $S$\\
$p, p_0, p_1,\dots$ 		&	Elements of the reference set $P$\\
$M$                     & A fixed radius used to define a partition with a given $p \in P$\\
$q,t$		&	A query $q \in U$ associated with a numeric query threshold $t$\\
$\mathcal{P}$			&	A partition of $S$ defined according to distances to $P$\\
$\mathcal{S}$			&	A class of $\mathcal{P}$ which may be excluded given a particular $q,t$\\
$\mathcal{Q}$			&	A subset of $U$ defined according to a particular $q,t$\\

$\mathbb{R}^n$	&	An $n$-dimensional real domain\\
$\ell_2$		& The Euclidean distance metric\\
$\tau$ 		&	A numeric parameter of the  Ptolemaic partitioning mechanism\\
&($\tau \ge 0.5$, typically $\tau \approx 1$)\\

\hline
\end{tabular}
\end{center}
\label{table_notation}
\end{table}%
It is generally assumed that $|S|$ is large and the cost of applying the function $d$ is high, and so the simple solution of applying $d(q,s)$ to all $s \in S$ is intractable \cite{zezula2006similarity}.

Table \ref{table_notation} gives a summary of these and other notations used thoughout the article.
\subsection{Filtering and Partitioning}

All metric search solutions rely upon algebraic properties of $(U,d)$. A relatively small set of distinguished reference points $P = \{p_0 ,\dots, p_m\} $ (typically, $P \subset S$) is used to avoid direct calculation of $d(q,s)$, after the distances $d(s,P)$ and $d(q,P)$ have been calculated. $d(s,P)$ is calculated ahead of query time, during a pre-processing phase. Two types of usage are distinguished as follows:

\begin{description}
\item[ filtering:] given a query $q \in U$, a specific datum $s \in S$, and the distances  $d(q,P)$ and $d(s,P)$, it may be possible to determine that  $d(q,s) > t$ for some $t$ without having to calculate $d(q,s)$.
\item[partitioning:] given a partition $\mathcal{P}$ of $S$ determined at pre-processing time with respect to $d(S,P)$, and the distances $d(q,P)$, it may be possible to determine that some classes of $\mathcal{P}$ do not contain any elements $s$ such that $d(q,s) \le t$.
\end{description}

Both types of mechanism have their place in metric search, see \cite{chavez_searching_2001_2,zezula2006similarity} for many examples. Filtering approaches however imply linear-time solutions, whereas partitioning can be used to construct an indexing mechanism, typically where a very large data set is recursively partitioned, in order to achieve a sub-linear search time.

For filtering, the algebraic properties are required to give a lower-bound on the distance $d(q,s)$ with reference to the sets of distances $d(q,P)$ and $d(s,P)$. For partitioning, a further requirement is to identify a partition that can be determined at pre-processing time, of which  one or more classes may be excluded at query time according to $d(q,P)$.

Table \ref{table_partition_mechanisms} shows partitioning mechanisms which derive from  various known lower-bound properties. The contribution of this paper is a novel partitioning mechanism for Ptolemaic inequality,  shown in bold type in the table. Until now, such a partitioning mechanism has been missing from the literature.

\begin{table}[]
\caption{Five different partition functions and their corresponding exclusion conditions.
In all cases the Partition Criterion is used to form a distinguished subset of $S$ at pre-processing time, for all $s \in S$. The Exclusion Condition is evaluated with respect to the query $q$ and a query radius $t$.
 Row 5 summarises the novel contribution of this paper.}
\begin{center}
\begin{tabular}{|l|l|c|c|}
\hline
&Underlying property
&Partition Criterion
& Exclusion Condition\\
\hline
1&
triangle inequality
& $d(s,p) \le M$
& $d(q,p) > M + t$
\\
\hline
2&triangle inequality
& $d(s,p) \ge M$
& $d(q,p)  < M - t$
\\
\hline
3&triangle inequality
& $d(s,p_0) \le d(s,p_1)$
& $d(q,p_0) - d(q,p_1) > 2t$
\\
\hline
4&four-point lower bound
& $d(s,p_0) \le d(s,p_1)$
& $\frac{d(q,p_0)^2 - d(q,p_1)^2}{d(p_0,p_1)} >2t$
\\
\hline
5&\textbf{Ptolemaic inequality}
& { \boldmath $d(s,p_0) \le d(s,p_1)$ }
& {\boldmath $d(q,p_0) - d(q,p_1) > t / \tau$}
\\ && { \boldmath $ \land \,\, d(s,p_1) \ge \tau d(p_0,p_1)$} &
\\
\hline
\end{tabular}
\end{center}
\label{table_partition_mechanisms}
\end{table}%
The remainder of this section introduces some necessary preliminaries.
In Section \ref{sec_analysis}  the underlying geometry of the partition mechanism is given, and Section \ref{sec_definition} gives its formal definition. Section \ref{sec_evaluation}  gives a quantitative analysis of its  value.

\subsection{Subclasses of Metric Space}

Properties (1-3) listed in Table \ref{table_partition_mechanisms} are possessed by all proper metric spaces. Property (4) is found only in \emph{supermetric} spaces \cite{Connor2016:SISAP_Supermetric}, which include all spaces which are isometrically embeddable in Hilbert space%
\footnote{see eg \url{https://en.wikipedia.org/wiki/Hilbert_space}},
while property (5) is found in any Hadamard space%
\footnote{see eg \url{https://en.wikipedia.org/wiki/Hadamard_space}}.

Any  Hilbert-embeddable  space is also a Hadamard space; although Hadamard spaces are a little more general, it is not clear that any practical non-Hilbert spaces fall in this category. Details of Hilbert spaces  are elaborated in \cite{Connor2016:HilbertExclusion}; in this context it is  sufficient to know that the following classes of metric space are members of both classes: Euclidean, Cosine, Jensen-Shannon, Quadratic Form, Triangular, and Mahalanobis spaces. Furthermore, the square root of any proper metric gives a space in both classes. The partition mechanism described here is thus applicable to any of these spaces.

\subsection{Ptolemaic and Four-point Lower Bounds}
\label{subsec_4pt_proj}

The Ptolemaic inequality was  identified for use as a distance lower-bound for certain  metric spaces in \cite{hetland_orig}, and used further in a number of studies for example \cite{HETLAND2013989,piOfSqfd}.
For any four objects  $u_0,u_1,u_2,u_3 \in U$, the Ptolemaic inequality states:
\[d(u_0,u_2) \cdot d(u_1,u_3) \le d(u_0,u_1) \cdot d(u_2,u_3) + d(u_1,u_2) \cdot d(u_3,u_0) \] 
In $(\mathbb{R}^n,\ell_2)$ this is more simply stated as the product of the lengths of the diagonals of any quadrilateral being no greater than the sum of the products of the pairs of opposing sides. Given this property, a lower bound on the distance $d(q,s)$ can be determined whenever, for two reference values $p_0,p_1$, all the distances $d(s,p_0), d(s,p_1),d(q,p_0), d(q,p_1)$ and $d(p_0,p_1)$ are known. This lower bound is much tighter than those available via simple triangle inequality, and has been used to great effect for filtering objects during search, particularly in the context of a very expensive distance function \cite{piOfSqfd}. The mechanisms used to incorporate this lower bound into metric search techniques include the Ptolemaic pivot table, the Ptolemaic PM-Tree, and the Ptolemaic M-Index \cite{HETLAND2013989}. In all cases, the inequality is used as an extra filtering mechanism superimposed onto an existing filtering or partitioning structure.


The four-point lower bound property, and the Hilbert exclusion mechanism, were first identified in \cite{Connor2016:SISAP_Supermetric}, and investigated further in \cite{Connor2016:HilbertExclusion,connor2019supermetric}.
Any supermetric space $(U,d)$ has the four-point property: for any four objects  $u_0,u_1,u_2,u_3 \in U$, there exists a tetrahedron with vertices $u_0',u_1',u_2',u_3' \in \mathbb{R}^3$ where the distances between pairs of points are preserved, i.e. $d(u_i,u_j) = \ell_2(u_i',u_j')$.

 The four-point property thus  implies the Ptolemaic property, but not vice-versa.
 
The four-point  lower-bound property  applies to the case where five of the six edge lengths of a tetrahedron are known. In this case, two adjacent faces of the tetrahedron can be constructed.
A lower bound of the final side length is obtained by notionally rotating these faces around their common edge to minimise the final edge length, which occurs when a planar tetrahedron is formed. 

%



\subsection{Projection into 2-dimensional Space}

\begin{figure}[] 
\begin{center}
\includegraphics[width=7cm]{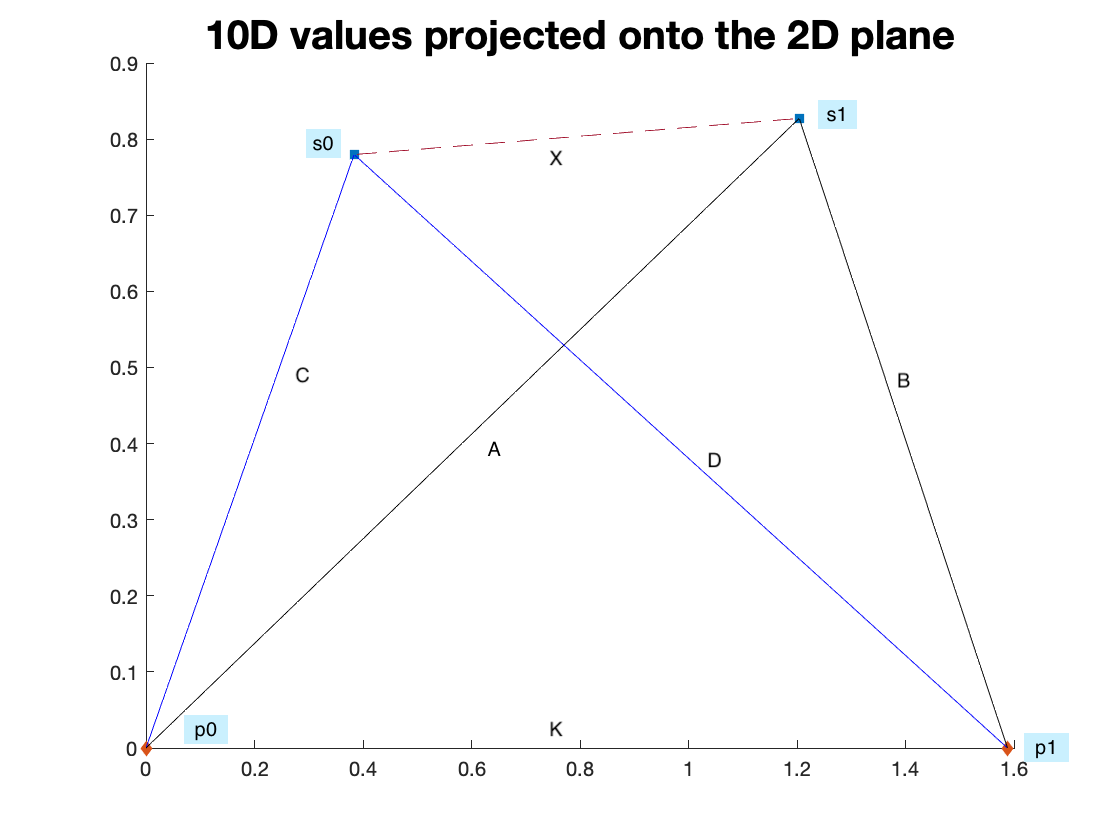}
\caption{Four objects  $p_0,p_1,s_0,s_1$ selected from a supermetric space are projected onto a 2D plane according to the known distances $K,A,B,C,D$. Although X is not known, it is known that a tetrahedron with these four vertices exists in 3 dimensions. By the four-point lower bound, $ d(s_0,s_1) \ge X$. By the Ptolemaic lower bound in 2 dimensions, $X \ge \frac{AD - BC}{K}$}
\label{fig_four_point_quad}
\end{center}
\end{figure}

Together these properties imply that if the Ptolemaic inequality is applied to a quadrilateral in two dimensions, when that quadrilateral has been formed according to  five distances measured among  four objects in a supermetric space, then the inequality applies also to the original space.  Figure \ref{fig_four_point_quad} shows an example of this.

The figure shows a projection in $(\mathbb{R}^2,\ell_2)$ of four objects $p_0,p_1,q$ and $s$ selected from a supermetric space $(U,d)$. All distances other than $d(q,s)$ have been calculated  in $(U,d)$.
The projections of objects $p_0$ and $p_1$ are plotted at the points $(0,0)$ and $(0,d(p_0,p_1))$ respectively%
\footnote{This choice is arbitrary, any two points which preserve $d(p_0,p_1)$ could be used}.
 The projections of objects $q$ and $s$ are plotted at the unique points above the X-axis which preserve their distances from $p_0$ and $p_1$. The supermetric properties imply that the tetrahedron $p_0,p_1,q,s$ must exist in $(\mathbb{R}^3,\ell_2)$, therefore the unknown distance $d(q,s)$ is lower-bounded by the sixth edge of the planar tetrahedron plotted in $(\mathbb{R}^2,\ell_2)$.

By the four-point lower bound property, $d(q,s) \ge X$. By the Ptolemaic lower-bound property, $X \ge \frac{AD - BC}{K}$. Therefore, in the original supermetric space, $d(q,s) \ge \frac{AD - BC}{K}$. For the rest of this article, only  2D projections like these are considered, where two distinguished reference objects $p_0,p_1$ are used to form a planar projection of the rest of the data set, and rely on the Ptolemaic property with the context of planar quadrilaterals. This  restricts the outcome to Hilbert-embeddable spaces, although as noted this is not a significant practical restriction.

It is worth noting that  while the derivation and correctness of the mechanism  rely upon the existence of the 2D projection, the  projection itself does not require to be calculated. As shown in Table \ref{table_partition_mechanisms}, the calculations required are restricted to  simple calculations over distances measured in the original space.

\section{The Underlying Geometry}

\label{sec_analysis}

Partitioning  mechanisms differ from filtering in that, for each possibility of exclusion, it is necessary to identify two subsets of the universal space:

\begin{enumerate}
\item a \emph{static} subset $\mathcal{S}$, which can be identified and indexed  during the pre-processing of the finite data set, and
\item a \emph{dynamic} subset $\mathcal{Q}$, which is identified only after the query and (typically)  its associated search radius become apparent.
\end{enumerate}

Exclusion of $\mathcal{S}$ can be  performed when every element of $\mathcal{Q}$ is separated by at least the search radius from every element of $\mathcal{S}$. In the following section objects denoted by $s$ and $q$ are referred to, representing elements of $\mathcal{S}$ and $\mathcal{Q}$ respectively.

\subsection{2D Geometry}
\label{sec_geometry}
\begin{figure}[]
\begin{center}

\begin{subfigure}[b]{0.4\textwidth}
	\includegraphics[width=\textwidth]{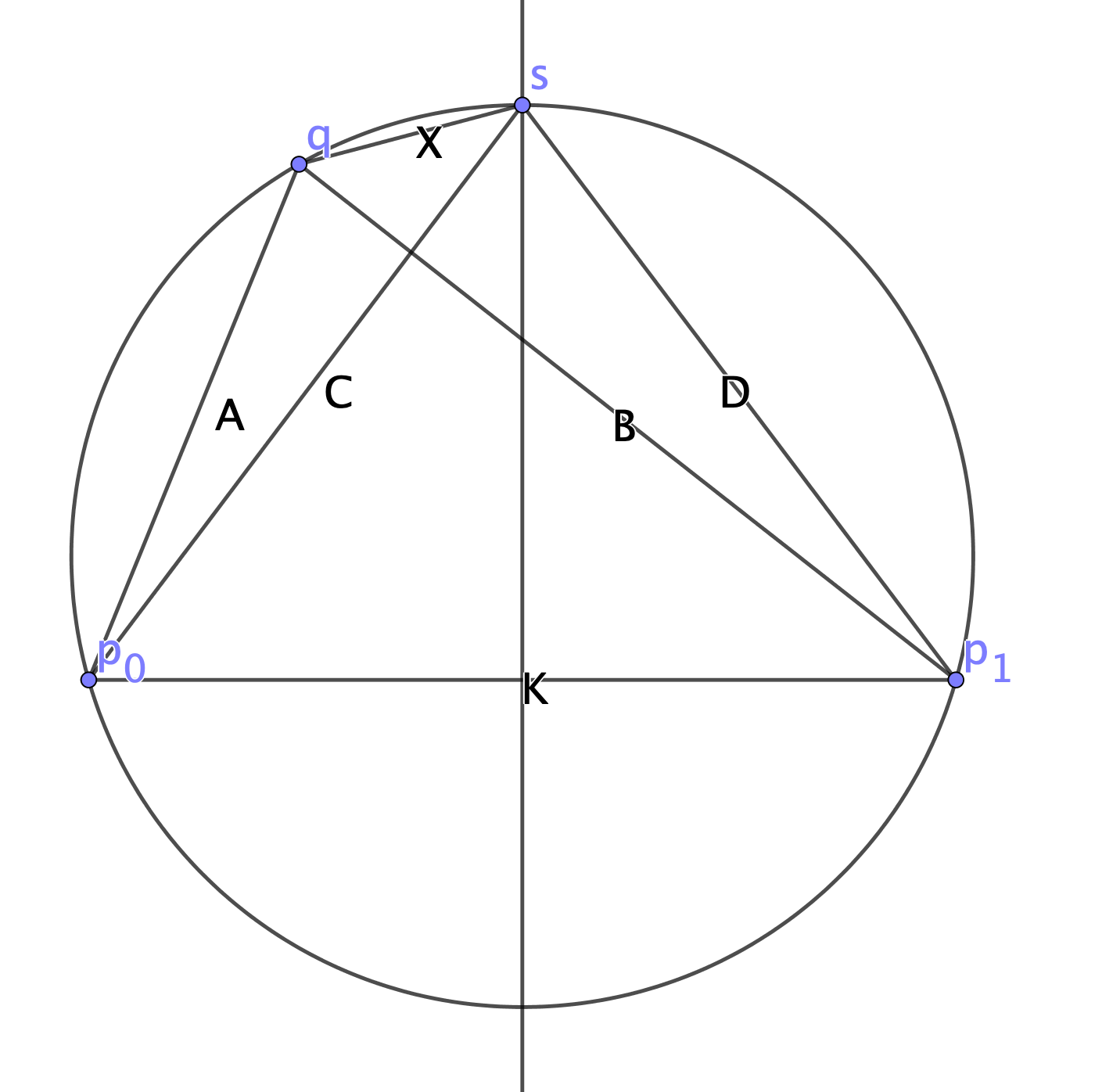}
	\caption{Four points $p_0,p_1,s,q$  on a 2D plane.}
	\label{fig_basic_quad_a}
\end{subfigure} \hfill
\begin{subfigure}[b]{0.5\textwidth}
	\includegraphics[width=\textwidth]{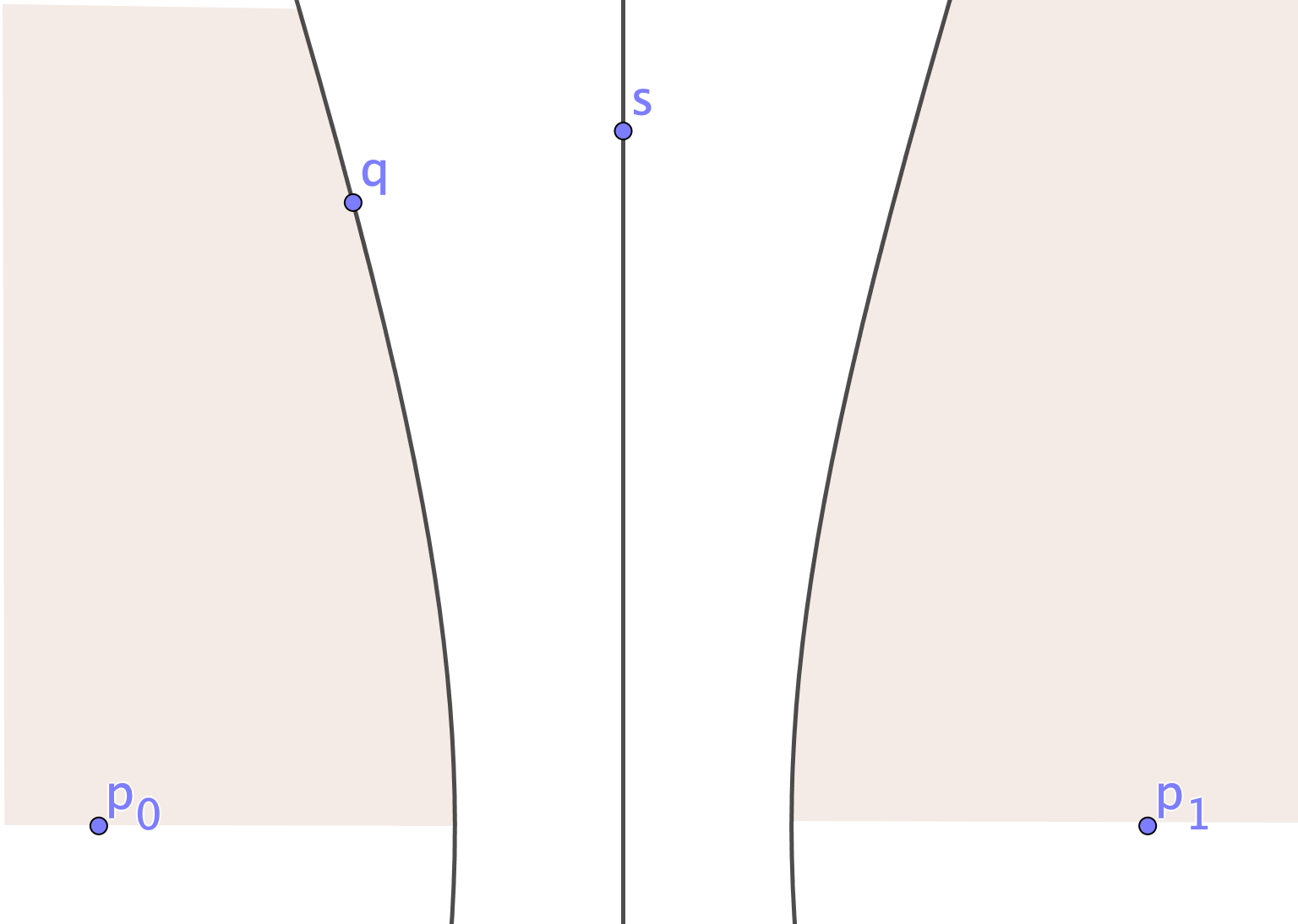}
	\caption{For any $q'$ in the shaded area, $d(q',s) \ge \tau|A-B|$.}
	\label{fig_basic_quad_b}
\end{subfigure}

\caption{ In \ref{fig_basic_quad_a}, line segments among the points are annotated by their lengths. $s$ is chosen according to the parameter $\tau = C / K$, and $q$ is a point on the circle defined by $p_0,p_1$ and $s$. $q$ is the unique point with $X = \tau(B-A)$.
The regions shown in the figure represent  regions of the original, potentially non-Euclidean, space;  the inequalities established are generally applicable to the original space.}
\label{fig_basic_quad}
\end{center}
\end{figure}

Figure \ref{fig_basic_quad_a} shows four points $p_0, p_1, q$ and $s$ drawn on a plane. 
 These points represent the 2D projections of two reference (or pivot) values $p_0,p_1$, a query value $q$ and a potential solution value $s$. The figure is annotated with line segments labelled $A - D$, $K$ and $X$, where the labels represent the lengths of the respective lines. $K$ is the  inter-pivot distance, and  $X$ is a lower bound of the unknown distance $d(q,s)$.
 
For the moment, values have been chosen such that 
 \begin{itemize}
\item $C = D$
\item the  parameter $\tau$ defines the  ratio $C / K$ 
\item point $q$  lies on the same circle as $p_0,p_1$ and $s$
 \end{itemize}

%

The Ptolemaic inequality  states
\[BC \le AD + KX\]
so in this  case:
\[X \ge  \tau(B-A)\]
%

The boundary of this region defines a hyperbola  with foci $p_0,p_1$ and semi-major axis $X/2\tau$, as shown in Figure  \ref{fig_basic_quad_b}. It follows that any point within the shaded region is at least distance $X$ from the point $s$.

\begin{figure}[]
\begin{center}
\begin{subfigure}[b]{0.45\textwidth}
	\includegraphics[width=\textwidth]{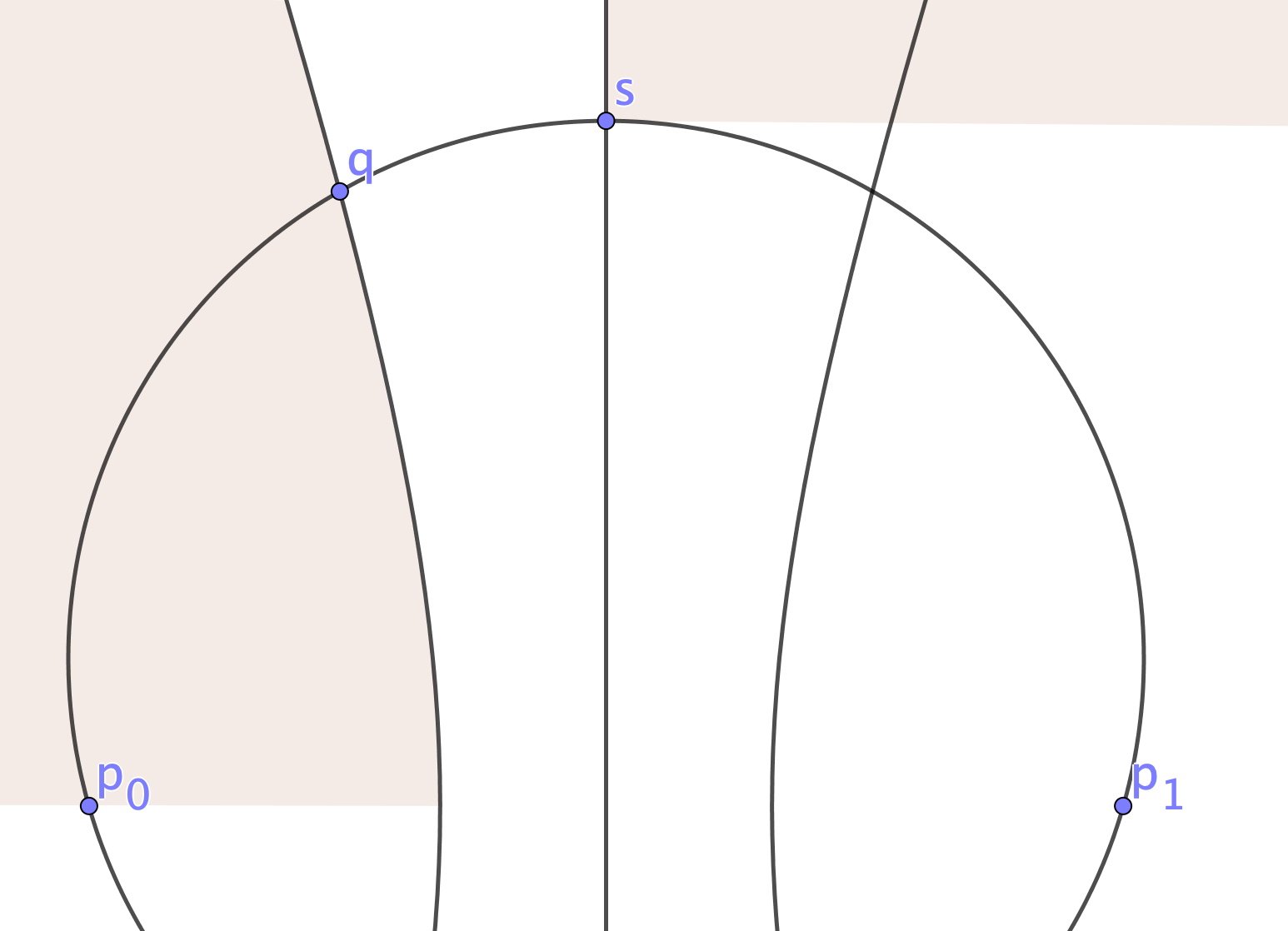}
         \caption{The shaded area to the right of the central axis contains points which are  at least $d(q,s)$ from that on the left. }
	\label{fig_static_partitions_a}
\end{subfigure}
     \hfill
\begin{subfigure}[b]{0.45\textwidth}
	\includegraphics[width=\textwidth]{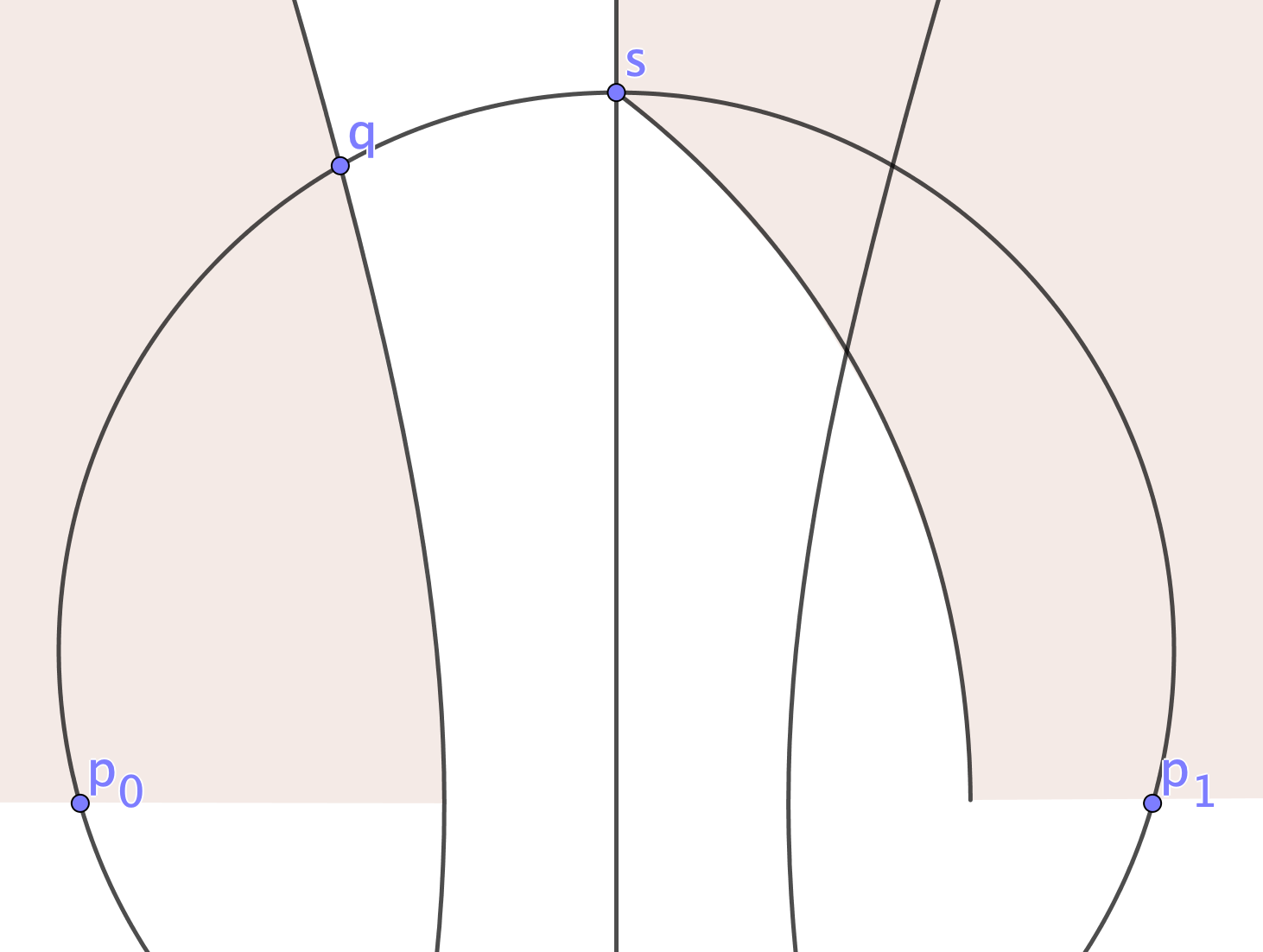}
         \caption{The shaded area here also contains contains points which are at least $d(q,s)$ from that on the left.}
	\label{fig_static_partitions_b}
\end{subfigure}
\caption{The shaded area to the left of each central axis denotes the locus defined by $d(q',p_0) - d(q',p_1) > \tau t$ for any $q' \in U$. In both cases, for any $s' \in S$ in the shaded area to the right of the central axis, $d(q',s') > t$.}
\label{fig_static_partitions}
\end{center}
\end{figure}

As $q, p_0,p_1$ and $s$ are co-circular, $q$ is  the  unique closest point on the (left-hand) hyperbola to $s$ and $X = \tau(B-A)$. The line segment $sq$ is therefore perpendicular to the tangent of the hyperbola at $q$. 
%
As the gradient of the tangent is  negative, it follows that any point above and to the right of $s$ is further than $X$ from any point to the left of the hyperbola, as illustrated in Figure \ref{fig_static_partitions_a}.  

%

However,  the static partition can be extended to include more of the finite search space, by including any value $s' \in S$ to the right of the central axis where also $d(s',p_0) \ge d(s,p_0)$, as illustrated in Figure \ref{fig_static_partitions_b}. This not only increases the cardinality of the potentially excluded subset, but also avoids the requirement to calculate the 2D projection.

The static and dynamic classes represented in Figure \ref{fig_static_partitions_b} are now formally defined as
\begin{align*}
\mathcal{S} \quad&= \quad \{s' \leftarrow S\, | \, d(s',p_0) \ge d(s',p_1) \quad\land\quad d(s',p_0) \ge \tau d(p_0,p_1)\} \\
\mathcal{Q} \quad&=\quad \{q' \leftarrow U\, | \, d(q',p_1) - d(q',p_0) > t/\tau \}
\end{align*}

with the property that it is impossible for any element $q' \in \mathcal{Q}$ to be within distance $t$ of any element  $s' \in \mathcal{S}$.
%

%
 The validity of this extension seems evident from the illustration in Figure \ref{fig_static_partitions_b}, but of course needs to be demonstrated for the general case.  A full justification of the correctness is included in Appendix \ref{appendix_a}.

\section{The Partition Mechanism}
\label{sec_definition}

\begin{figure}[]
\begin{center}
\begin{subfigure}[b]{0.45\textwidth}
	\includegraphics[width=\textwidth]{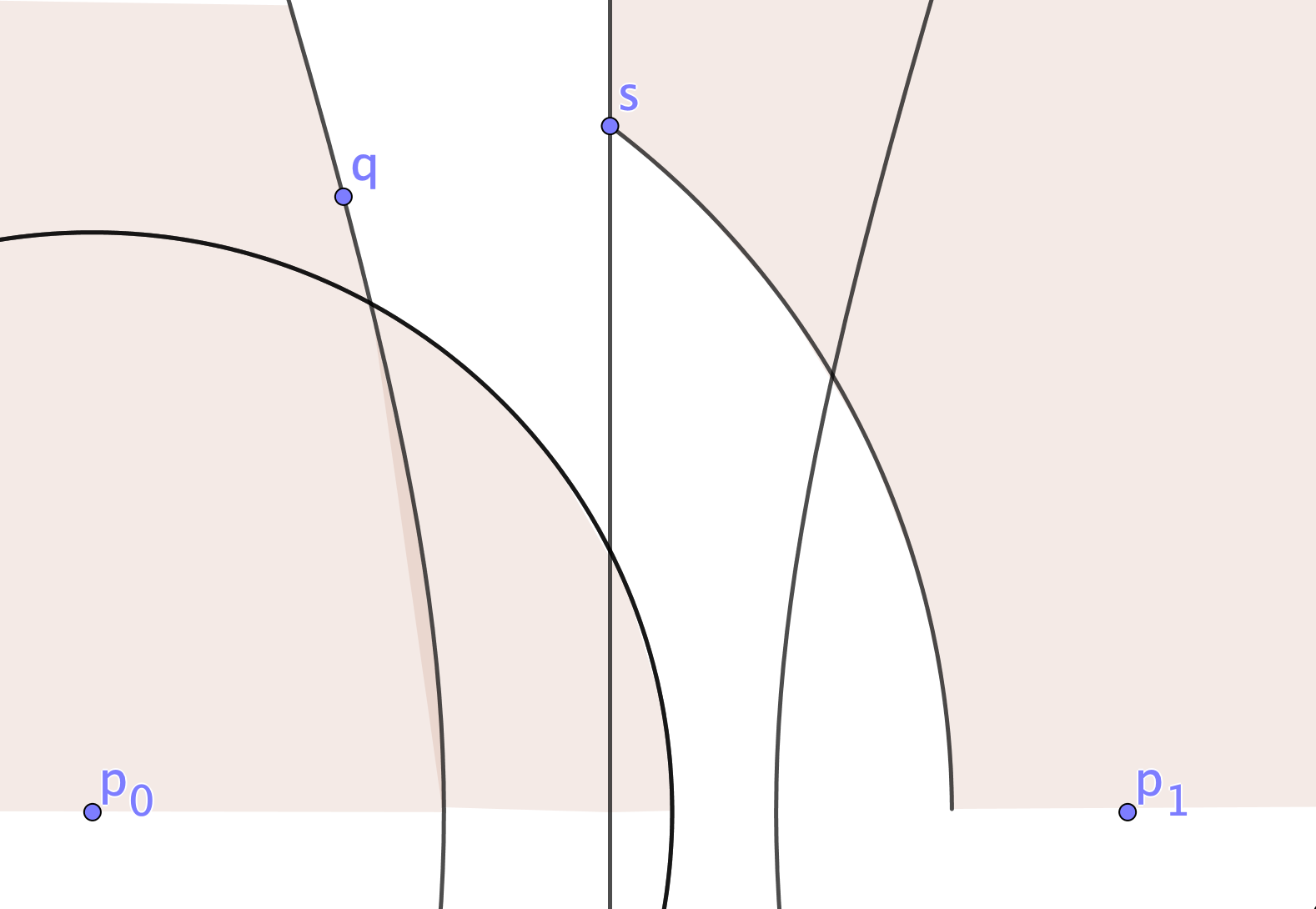}
         \caption{The locus $d(p_0,q) < d(p_0,s) - t$ can be included in $\mathcal{Q}$.}
	\label{fig_final_static_partition_a}
\end{subfigure} \hfill
\begin{subfigure}[b]{0.45\textwidth}
	\includegraphics[width=\textwidth]{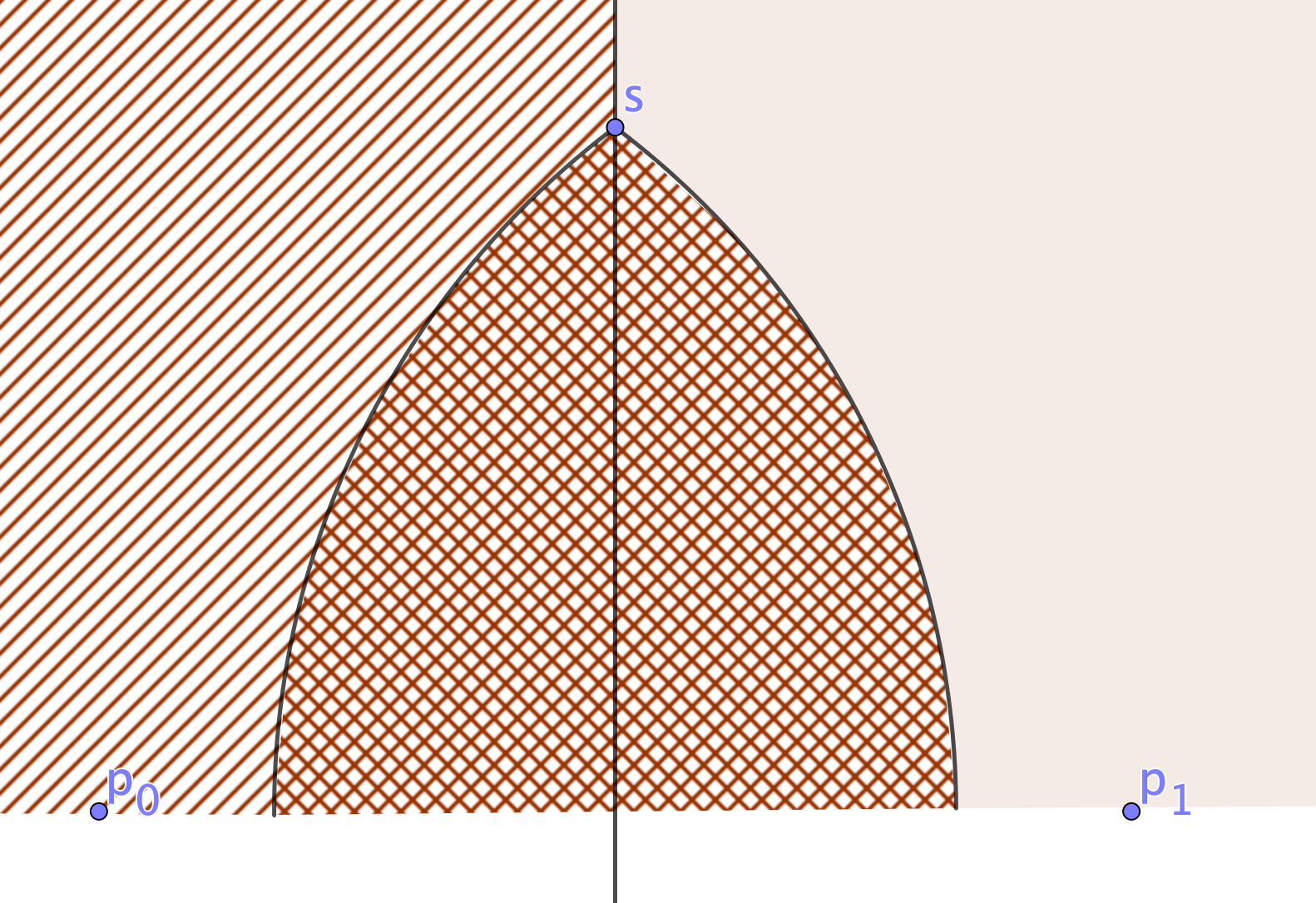}
         \caption{The locus $d(s,p_{\{0,1\}}) < \tau d(p_0,p_1)$ forms the final partition class.}
         \label{fig_final_static_partition_b}
\end{subfigure}
     \hfill
\caption{}
\label{fig_final_static_partition}
\end{center}
\end{figure}

The addition of the  criterion $d(s',p_0) \ge \tau d(p_0,p_1)$ to the static partition allows further exclusion potential relying on the normal triangle inequality method, i.e. if $d(q,p_0) < \tau d(p_0,p_1) - t$.
This extension to the exclusion criterion is illustrated in Figure \ref{fig_final_static_partition_a}.

Furthermore, when also including the symmetric opposite criteria, the static partition now defines  three subclasses as shown in Figure \ref{fig_final_static_partition_b}. It may further be noted that the third of these subclasses may also independently excluded if $d(q,p_0)$ or $d(q,p_1) \ge \tau K + t$,  again relying only on triangle inequality.

%

So finally, according to the geometry established in Section \ref{sec_geometry}, a  static partition $\{\mathcal{S}_1,\mathcal{S}_2,\mathcal{S}_3\}$   of $S$ can be established for a pair of reference points $p_0$ and $p_1$ with $K = d(p_0,p_1)$ and a given value of $\tau$ as follows:
\begin{align*}
\mathcal{S}_1 \quad&= \quad \{s \leftarrow S\, | \, d(s,p_0) \ge d(s,p_1) \quad\land\quad d(s,p_0) \ge \tau K\} \\
\mathcal{S}_2 \quad&= \quad \{s \leftarrow S\, | \, d(s,p_0) < d(s,p_1) \quad\land\quad d(s,p_1) \ge \tau K\} \\
\mathcal{S}_3 \quad&= \quad \{s \leftarrow S\, | \, d(s,p_0) < \tau K \quad\land\quad d(s,p_1) < \tau K\}
\end{align*}

These static regions are illustrated on the 2D plane in Figure \ref{fig_final_static_partition_b}.

For a given query object $q$ with threshold $t$, where $A = d(q,p_0)$ and $B=d(q,p_1)$, these  classes can be excluded from a search as follows:
\begin{align*}
\mathcal{S}_1 \quad&: \quad B-A > t/\tau \quad\lor\quad A < \tau K - t\\
\mathcal{S}_2 \quad&: \quad A-B \ge t/\tau \quad\lor\quad B < \tau K - t \\
\mathcal{S}_3 \quad&: \quad A \ge \tau K + t \quad\lor\quad B \ge \tau K + t
\end{align*}

Note that it is possible for the exclusion of region $\mathcal{S}_3$ to occur in conjunction with that of $\mathcal{S}_3$ or $\mathcal{S}_3$.
The mechanism resulting from these definitions is now evaluated in Section \ref{sec_evaluation}.

\section{Evaluation}
\label{sec_evaluation}

Before proceeding with a  full quantitative evaluation, it is interesting to view graphical representations based on a sample from a particular data set, in order to  give a  more pragmatic view of the  Ptolemaic partition mechanism in comparison with  hyperplane (hyperbolic) and Hilbert (four-point) partition mechanisms.

\begin{figure}[]
\begin{center}
\begin{subfigure}[b]{0.45\textwidth}
	\includegraphics[width=\textwidth]{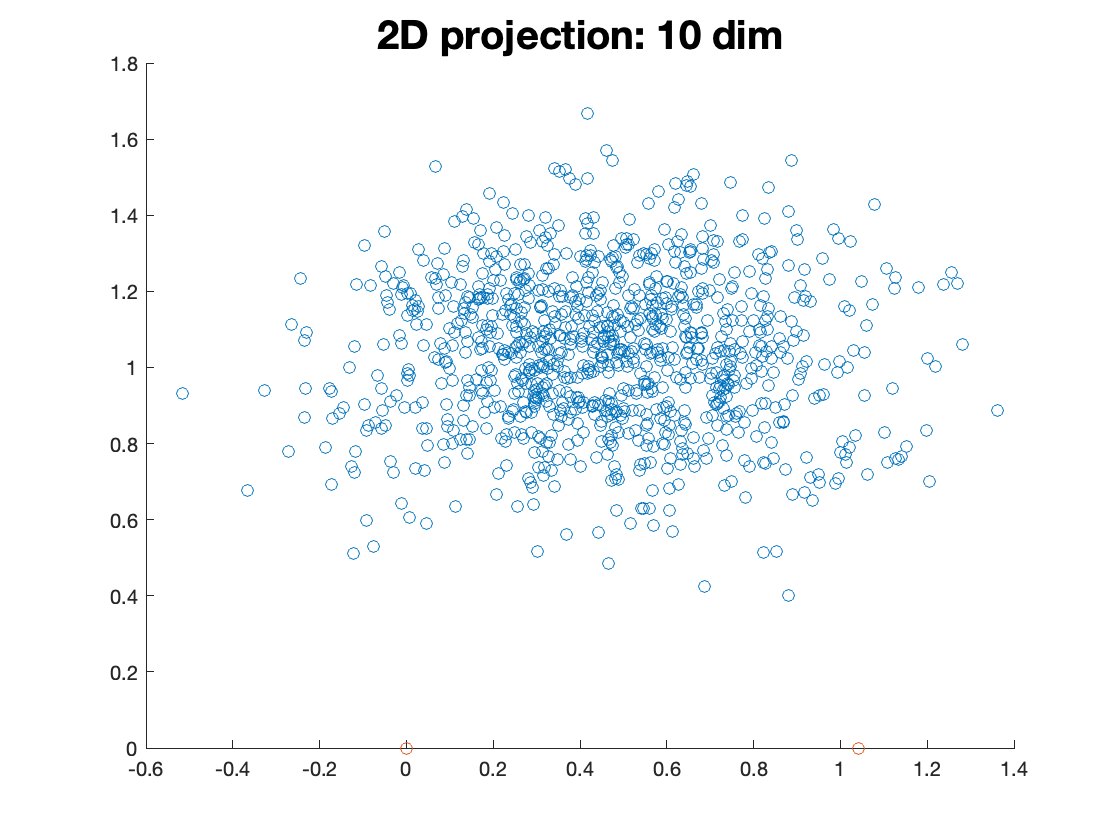}
	\caption{A scatter plot based on randomly selected reference points. $p_0$ is plotted at $(0,0)$ and $p_1$ at $(0,d(p_0,p_1))$. The data set is plotted according to the distance of each value from $p_0$ and $p_1$.}
	\label{fig_demos_1_a}
\end{subfigure} \hfill
\begin{subfigure}[b]{0.45\textwidth}
	\includegraphics[width=\textwidth]{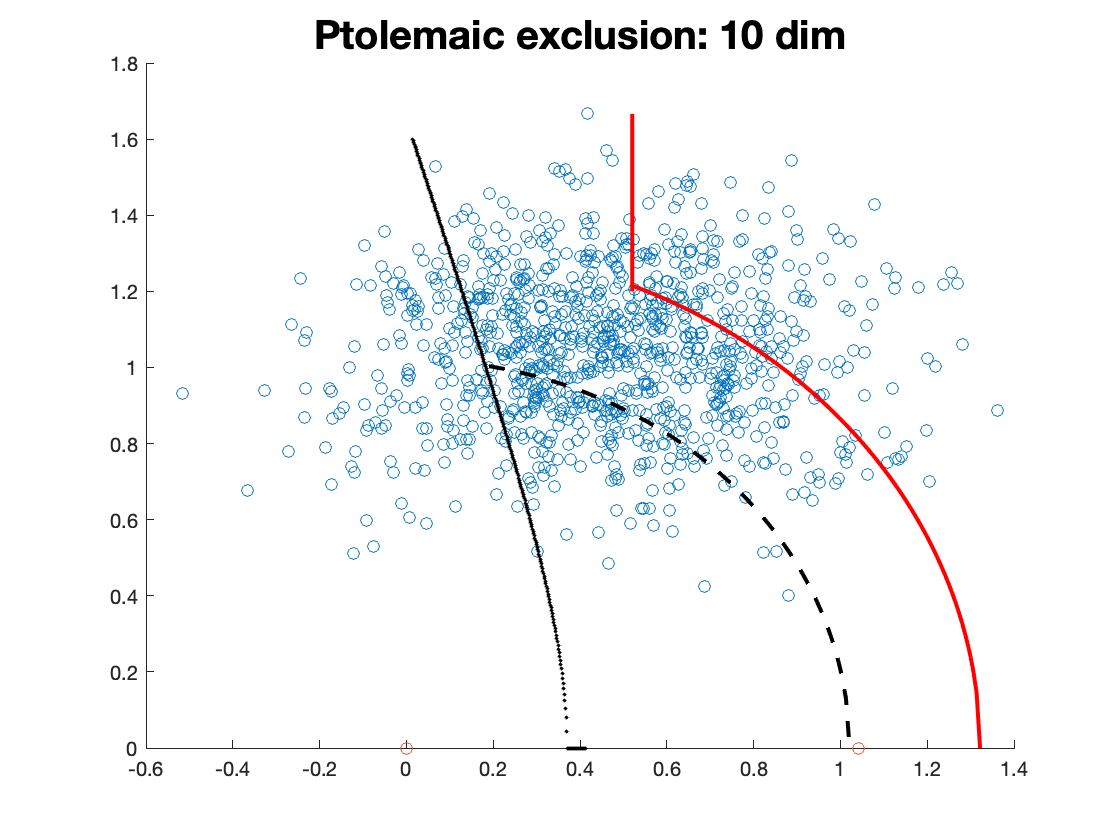}
	\caption{The boundary of the class $\mathcal{S}_1$ is plotted in red. Values lying to the right of this boundary can be excluded for queries lying to the left of either black boundary.}
	\label{fig_demos_1_b}
\end{subfigure}
\caption{Graphical view of the Ptolemaic partition mechanism based on a 2D projection.  A value of $\tau=1.3$ has been arbitrarily selected, along with a query threshold of 0.3, which is the mean nearest-neighbour distance within a set of 100k objects. }
\label{fig_demos_1}
\end{center}
\end{figure}

\begin{figure}[]
\begin{center}
\begin{subfigure}[b]{0.45\textwidth}
	\includegraphics[width=\textwidth]{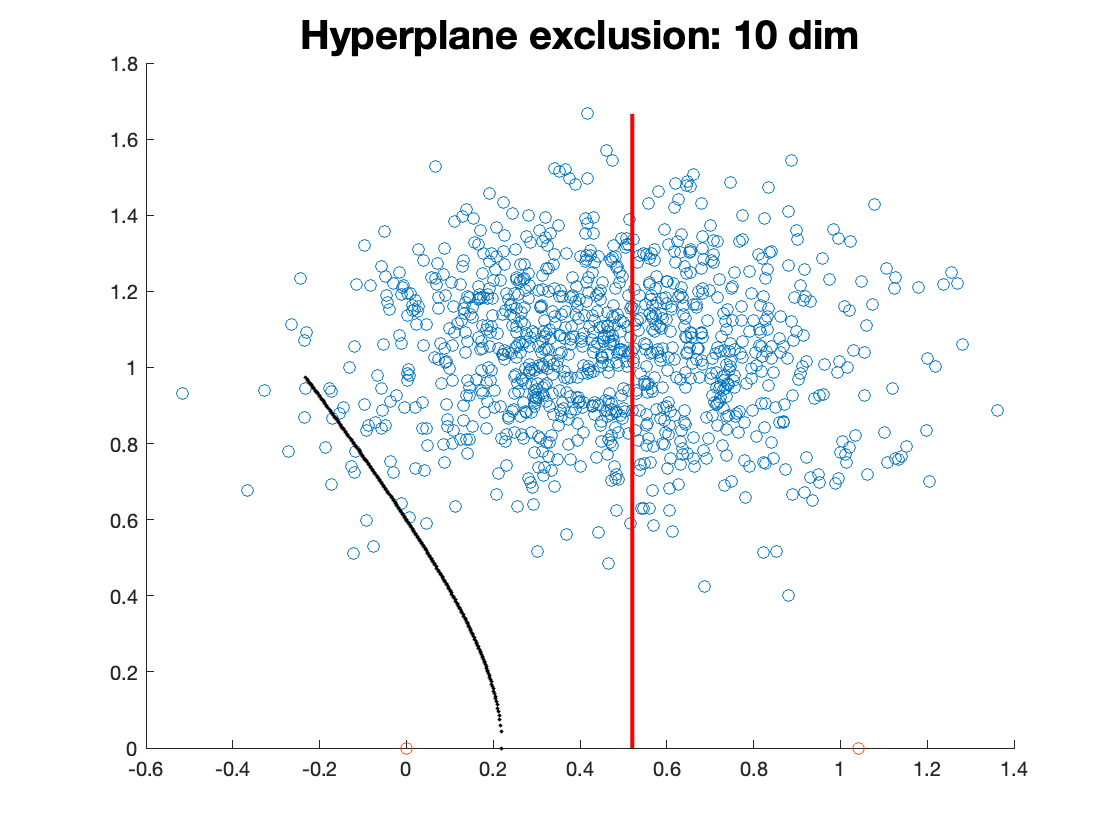}
	\caption{}
	\label{fig_demos_2_a}
\end{subfigure} \hfill
\begin{subfigure}[b]{0.45\textwidth}
	\includegraphics[width=\textwidth]{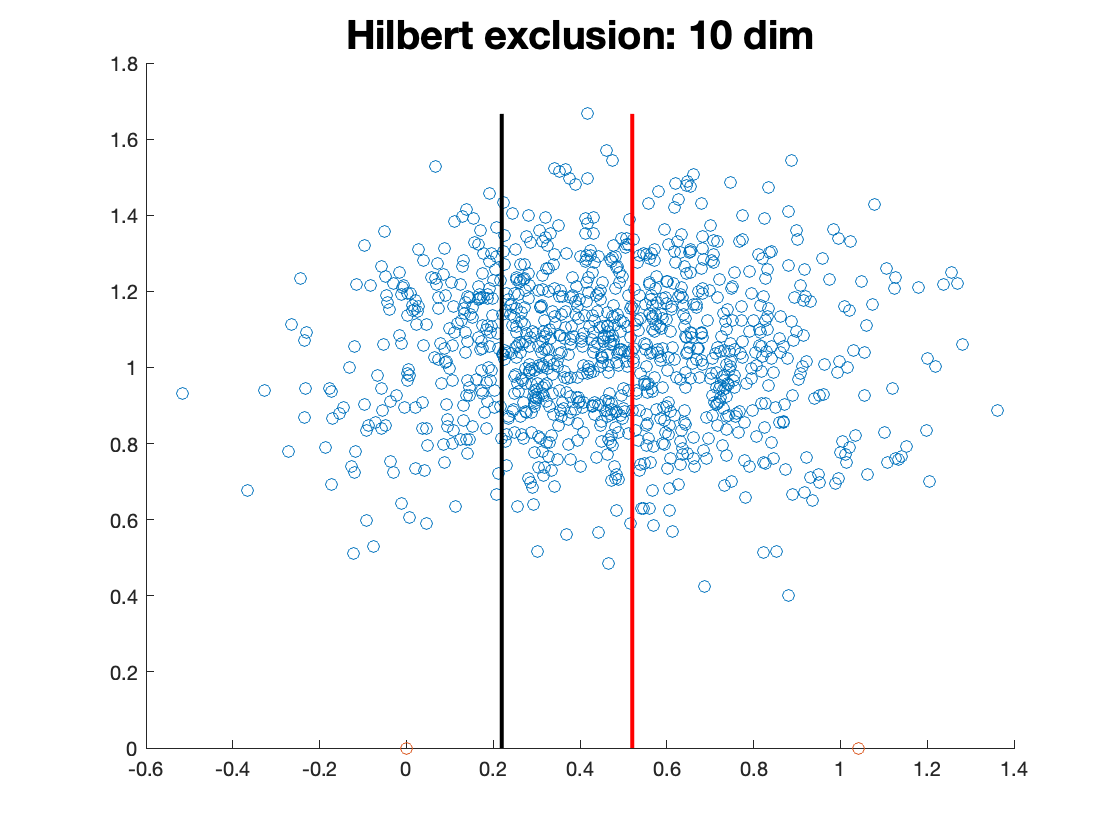}
	\caption{}
	\label{fig_demos_2_b}
\end{subfigure}
\caption{Equivalent graphical views of hyperplane and Hilbert partition mechanisms. In each case, queries falling to the left of the black line can be used to exclude the subset of data falling to the right of the red line. Again a query threshold of 0.3 has been used.}
\label{fig_demos_2}
\end{center}
\end{figure}

Figure \ref{fig_demos_1_a} shows a scatter plot of 1,000  generated values from a 10-dimensional Euclidean space, each projected onto a 2D plane according to their distances from two randomly generated pivot values. 

Figure \ref{fig_demos_1_b} shows the same projection superimposed with one of the partition boundaries of the Ptolemaic partition mechanism with a $\tau$ value of 1.3, and a query threshold of 0.3.
The boundary of the static region $\mathcal{S}_1$ is shown in red; those points lying to the right of the red boundary are thus subset to exclusion when either $d(q,p_1) - d(q,p_0) > 1.3 \cdot 0.3$, or if $d(q,p_1) < 1.3 \cdot d(p_0,p_1) - 0.3$. The boundaries of these regions are shown by solid and dotted black lines respectively; every point to the left of either boundary represents a value for $q' \in \mathcal{Q}$  which allows exclusion of the static class.

Figure \ref{fig_demos_2} shows the same plot with boundaries for standard hyperplane and Hilbert exclusion, in  \ref{fig_demos_2_a} and \ref{fig_demos_2_b} respectively. 

It is clear, at least in this case, that the Ptolemaic mechanism always excludes a smaller subset, while the probability of the exclusion being possible is  higher. From Figure \ref{fig_demos_2_a} it is evident that the dimensionality of the data set is starting to challenge hyperplane exclusion, while both Ptolemaic and Hilbert mechanisms continue to remain effective. Finally, while it is not possible to judge the relative efficacy of Ptolemaic vs. Hilbert from these diagrams, it can be observed that neither is a proper subset of the other, and it is therefore possible to use both Ptolemaic and Hilbert with respect to the same pair of reference points. This would allow a hybrid mechanism, more effective that either in isolation, based on the same dynamic measurements of $d(q,p_0)$ and $d(q,p_1)$.
\subsection{Quantitative Evaluation}

Quantitative evaluation is performed over sets of uniformly generated Euclidean data, from between 8 and 20 dimensions. 50k data objects are used and 1k non-intersecting queries are evaluated. The threshold used for each query corresponds to the 5nn distance as pre-calculated over the data.

Experiments were performed over Ptolemaic, Hilbert, and hyperplane mechanisms. For each experiment, a fixed number of reference points was used, and each of the $n \choose 2$ pairs of reference points was used to construct a partition over the space. The single outcome is the mean proportion per query of values that were successfully excluded, this value being between 0 and 1. For the majority of the experiments 10 reference points  used, this giving 45 different partitions. Thus all results  given correspond to the proportion of the data that can be successfully excluded at cost of only 10 distance calculations per query.

All experiments were performed using MatLab, and the code is available from the author%
\footnote{note to reviewers, will be published!}.
%

\subsection{Choosing $\tau$}

\begin{figure}[]
\begin{center}
\includegraphics[width=8cm]{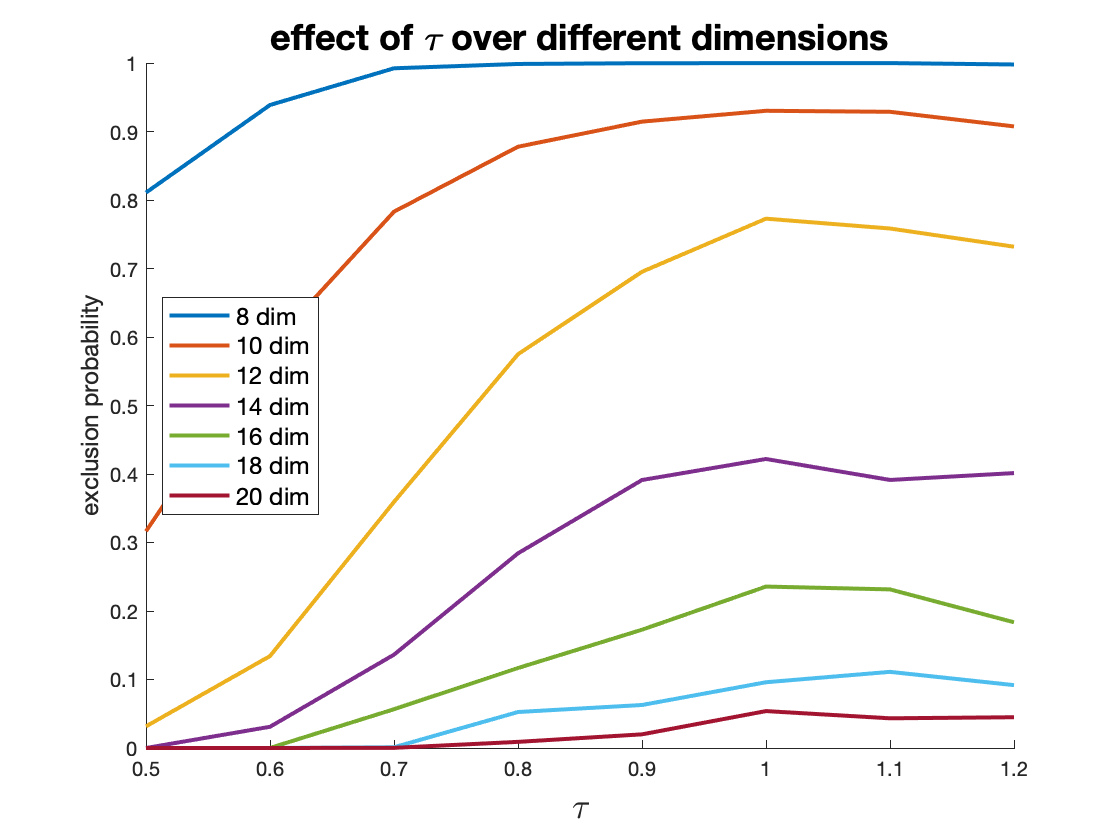}
\caption{Probability of successful exclusion for differing values of $\tau$ in different dimensions. Note that when $\tau=0.5$ the mechanism is identical to traditional hyperplane (Hyperbolic) exclusion.  }
\label{fig_eval_tau}
\end{center}
\end{figure}

First, different values for $\tau$ are examined. As mentioned, when $\tau = 0.5$ the mechanism reverts to simple hyperplane exclusion; while a value of less than $0.5$ is technically possible, there is no value in such a choice. As $\tau$ gets large, then ever fewer data will be present in the partition which may be excluded, and again the mechanism will become useless. Early tests showed that a value somewhere around $1$ is usually close to optimal, although for specific reference point pairs an optimum values of between around 0.8 and 1.2 were observed.

It would in fact be possible to optimise $\tau$ based on each particular pair of reference points, which we have not yet investigated thoroughly. In this experiment a  fixed  value of $\tau$ is  applied to all partitions, which is possibly more realistic for many scenarios.

Figure \ref{fig_eval_tau} shows the results of various values of $\tau$ when applied to data of between 8 and 20 dimensions. As can be seen there is a general trend of larger values being better as dimensions increase, but only within quite a small margin; while there is clearly an element of noise in this experiment, the best value in each case is either 1.0 or 1.1, although further investigation is warranted. For further experiments described over the data of different dimensions, the best value of $\tau$ found in this experiment was used.

\subsection{Evaluation over High Dimensional Data}

\begin{figure}[]
\begin{center}
\includegraphics[width=8cm]{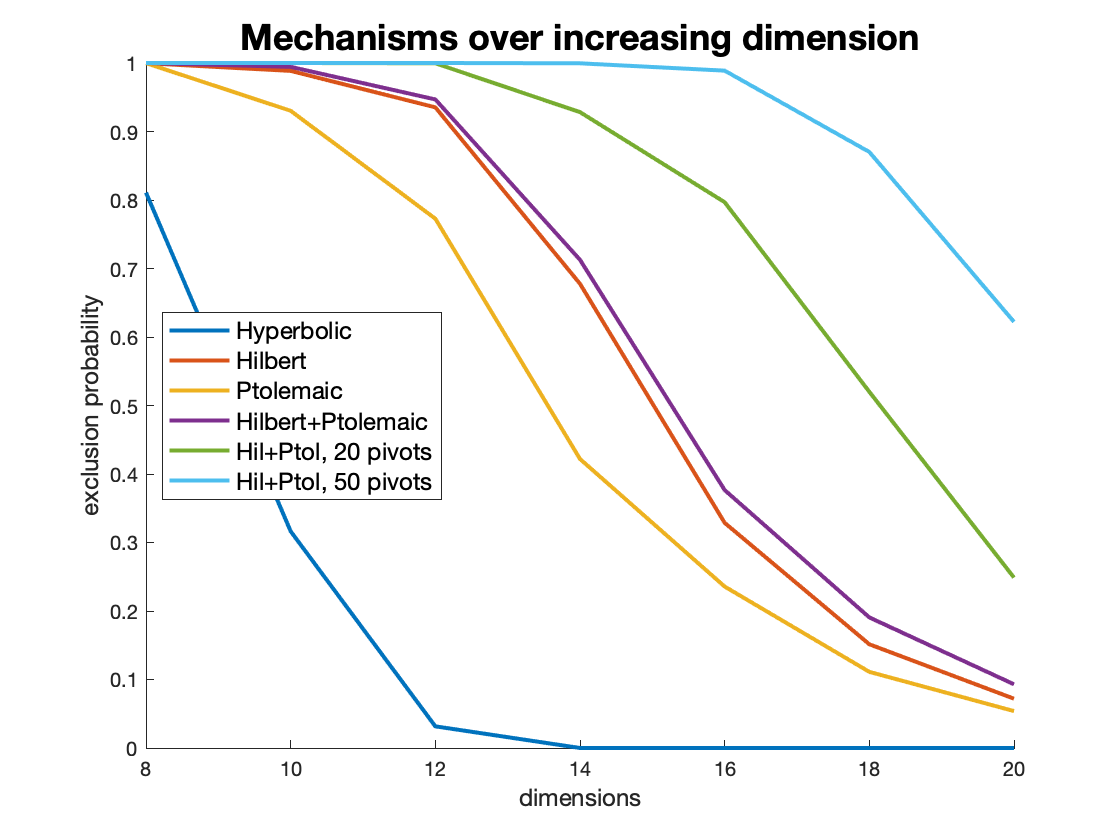}
\caption{ Performance of different exclusion mechanisms as dimensionality increases, measured as the probability of a non-solution being excluded based on the distances among query and reference values.}
\label{fig_eval_main}
\end{center}
\end{figure}

Having picked a value for $\tau$, outcomes for some different mechanisms  over  data ranging from 8 to 20 dimensions are given. Four mechanisms are used: Hyperplane (Hyperbolic) exclusion; Hilbert exclusion; Ptolemaic exclusion, and finally a combination of Hilbert and Ptolemaic exclusion. 

For the combination, for each pair of reference points five different subsets were identified at pre-processing time: two for Hilbert, and three for Ptolemaic, and all possible exclusions were attempted for each query. The observation here is that it is possible for different data to be excluded by each mechanism. As the essential query-time cost of performing the exclusions is the cost of the two distance operations and some relatively cheap arithmetic, taking the union of all possible exclusions makes practical sense as the distance calculations are amortised. As Hilbert exclusion always allows exclusion from a superset of queries identified by Hyperbolic exclusion, there is no point in combining that mechanism also.

Figure \ref{fig_eval_main} shows the outcome. As can be seen, while the performance of Hyperbolic exclusion falls rapidly away after around 8-10 dimensions, both Ptolemaic and Hilbert perform much better into the higher dimensional range. Hilbert always performs better than Ptolemaic, which is not very surprising as the four-point lower bound property is stronger then the Ptolemaic inequality, and technically applies to a smaller subset of metric spaces. What is more interesting, however, is that the combination of Ptolemaic and Hilbert gives a strictly better result than Hilbert alone; that is, the data sets identified for exclusion by the two mechanisms are not in a strict subset relation. Again it is noted that the inherent query-time cost of the joint mechanism is very similar to the cost of just one, as in all cases the query to pivot distances calculated are reused in both mechanisms.

The final plots in the graph show the use of 20 and 50 pivot values for the combination mechanism. Although only doubling the number of query-to-pivot distances required, 20 pivots gives $20 \choose 2$ i.e. 190 partitions to apply, and as can be seen the increase makes for a much higher exclusion ratio. Similarly, 50 pivots gives 1,225 partitions. The important observation however is that there is a clear degree of orthogonality in the randomly selected partitions, allowing almost perfect exclusion in 12 and 16 dimensions respectively.

\section{Conclusions and Future Work}

This paper fills a significant gap in the literature, that is a set partition that can be used as an exclusion mechanism for the Ptolemaic inequality; for some years, other distance lower-bounds have had known mechanisms and in this sense the Ptolemaic inequality has been an outlier.

In its simplest form, the mechanism is quantitatively much better than traditional hyperplane partitioning, and not quite as good as Hilbert partitioning. This is almost inevitable, as the class of spaces to which the inequalities can be applied are in a strict subset relation. Should this mechanism have been identified before Hilbert exclusion it would have been deserving of significant excitement, but this is nowadays tempered by the existence of the more effective Hilbert exclusion over essentially the same subclass of metric spaces.

However, it is the case that the individual data objects which the new mechanism excludes are not a proper subset of those identified by Hilbert exclusion, and as shown the two mechanisms may operate in conjunction to give a unified mechanism which, for the same cost of distance calculations against  reference points, gives a better exclusion outcome than either in isolation. Particularly in high-dimensional spaces, this therefore gives a further increment in the limit of dimensionality for which exact search can be effective. While the ``rule of thumb'' used to be that 8-10 dimensions was the effective limit for exact search \cite{Weber1998}, with the combined mechanism 16 dimensions can be effectively searched while avoiding almost all explicit distance calculations.

Some further avenues are worth exploring. First, it is feasible to calculate individual $\tau$ values customised to each particular pair of pivot points, rather than to choose a single value for the whole set. This would be expected to  give significant, if incremental, improvement in performance.

Finally, there are many other contexts beyond a simple recursive decomposition of a large data set where such mechanisms can be used. It is therefore of potential value in its own right for this previous gap in knowledge to be filled.

\section{Acknowledgements}
The author would like to sincerely thank the anonymous reviewers for their thorough and helpful comments on the submitted version of this article.

\bibliographystyle{splncs04}
\bibliography{bib} 

\section*{Appendix}
\appendix
\section{Justification of Correctness}
\label{appendix_a}

\begin{figure}[]
\begin{center}    
	\includegraphics[width=0.5\textwidth]{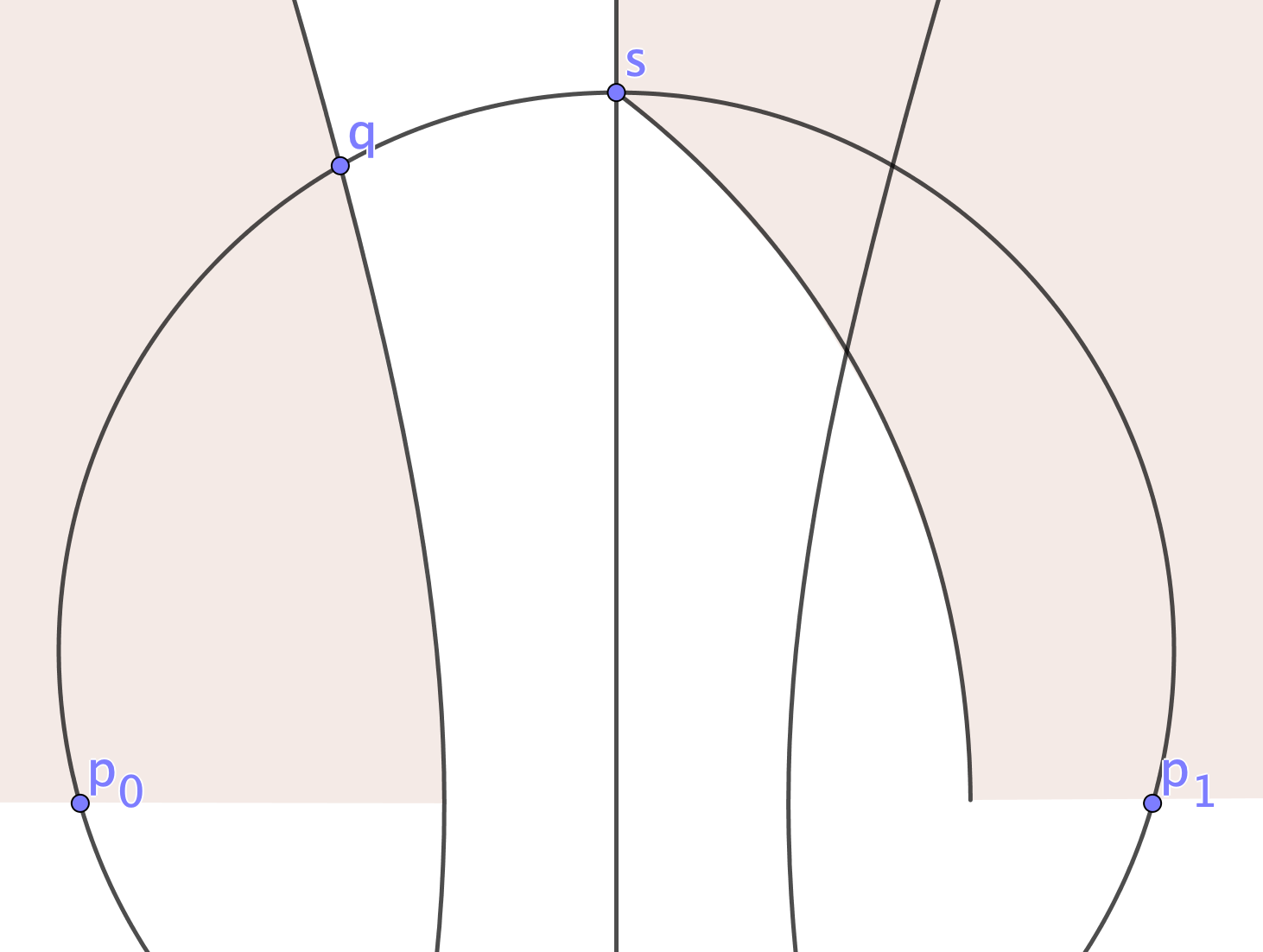}
\caption{The shaded area to the left of each central axis denotes the locus defined by $d(q',p_0) - d(q',p_1) > \tau t$ for any $q' \in U$. In both cases, for any $s' \in S$ in the shaded area to the right of the central axis, $d(q',s') > t$.}
\label{app_fig_static_partitions}
\end{center}
\end{figure}

Figure \ref{app_fig_static_partitions} is reproduced from the  text of Section \ref{sec_definition}. In the following  we  define the shaded regions  as:
\begin{align*}
\mathcal{S} \quad&= \quad \{s' \leftarrow S\, | \, d(s',p_0) \ge d(s',p_1) \quad\land\quad d(s',p_0) \ge \tau d(p_0,p_1)\} \\
\mathcal{Q} \quad&=\quad \{q' \leftarrow U\, | \, d(q',p_1) - d(q',p_0) > t/\tau \}
\end{align*}

 We now show that for any $q' \in \mathcal{Q}$ and $s' \in \mathcal{S}, d(q',s') > d(q,s)$%
 \footnote{Note that $s \in \mathcal{S}$ but $q \notin \mathcal{Q}$.}.
 
\begin{figure}[]
\begin{center}
\begin{subfigure}[b]{0.45\textwidth}
	\includegraphics[width=\textwidth]{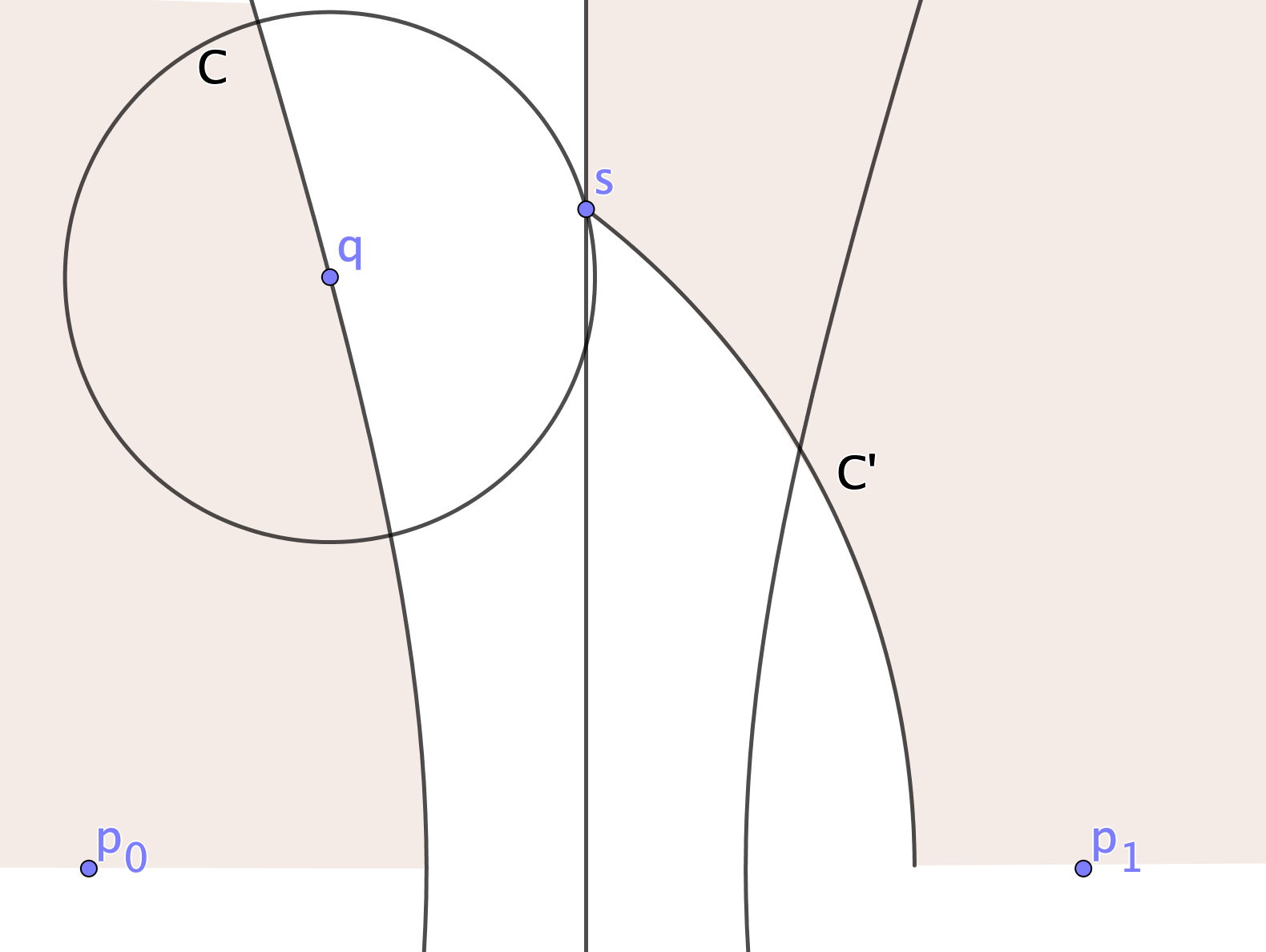}
         \caption{Circle $\mathcal{C}$ and arc $\mathcal{C'}$ must not intersect when $\mathcal{C}$ is centred at or below $q$.}
	\label{fig_gradient_proof_a}
\end{subfigure}
     \hfill
\begin{subfigure}[b]{0.45\textwidth}
	\includegraphics[width=\textwidth]{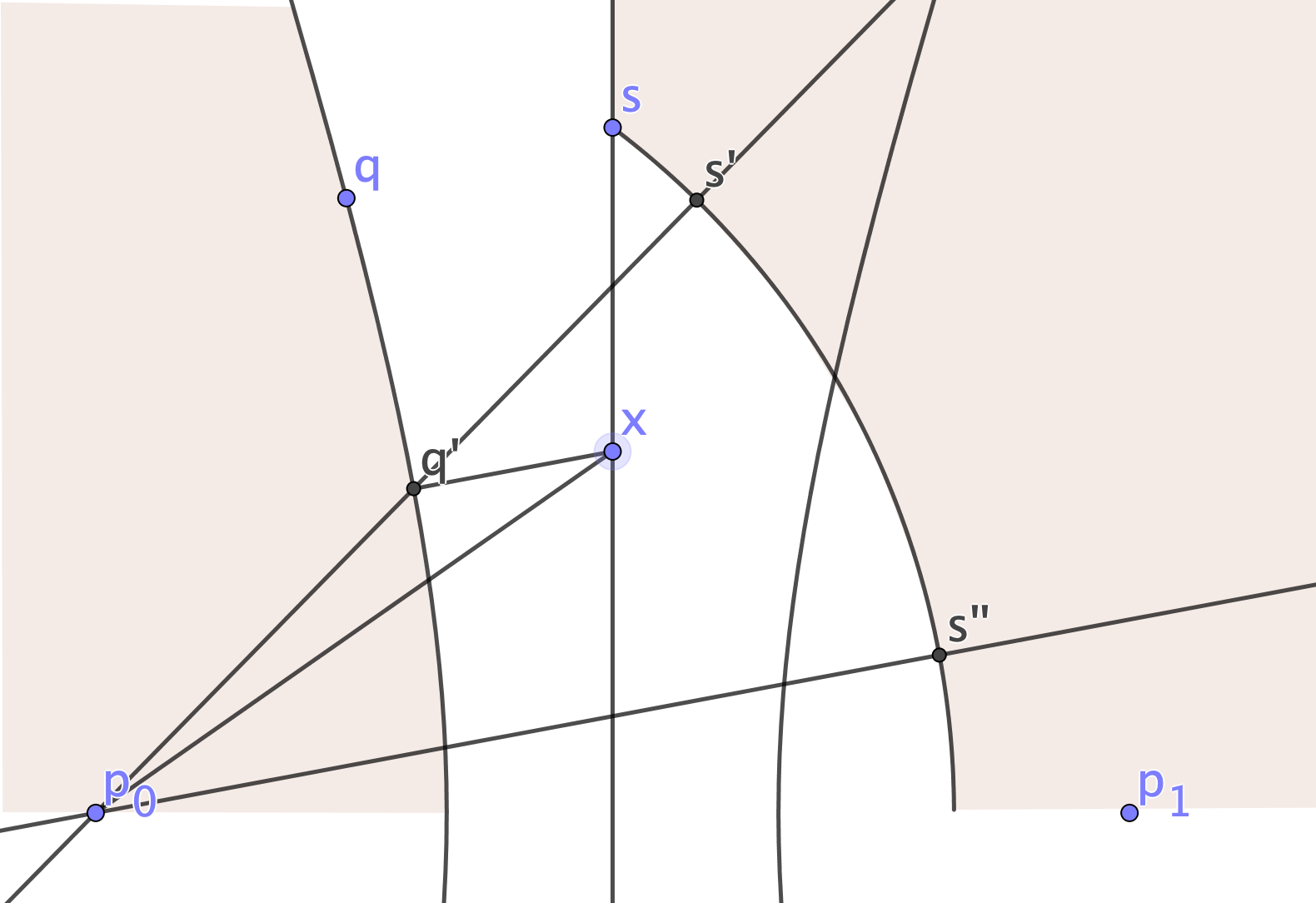}
         \caption{The gradient of the hyperbola at $q'$ is equal to the gradient of $\mathcal{C'}$ at $s''$.}
         \label{fig_gradient_proof_b}
\end{subfigure}
\caption{}
\label{fig_gradient_proof_1}
\end{center}
\end{figure}

 There are two circles to consider, illustrated in Figure \ref{fig_gradient_proof_a}. $\mathcal{C}$ is centred around a point on the hyperbola with radius $d(q,s)$, and $\mathcal{C}'$ is centred around $p_0$ with radius $d(p_0,s)$. So far we have shown  that $q$ is the nearest point on the hyperbola to $s$, and therefore $s$ lies on $\mathcal{C}$ when it is centred at $q$.  It is sufficient to show that $\mathcal{C}$ does not intersect with the boundary of $\mathcal{S}$ when its centre is at any point on the hyperbola containing $q$.

If $\mathcal{C}$ is centred around $q$, as illustrated in Figure \ref{fig_gradient_proof_a}, it  crosses the central axis. However as $p_0,q,s,$ and $p_1$ are co-circular, then (a) its radius is smaller than that of $\mathcal{C}'$, and (b) line $p_0s$ has a steeper gradient that line $qs$. Together these ensure that no element of $\mathcal{S}$ is within distance $d(q,s)$ of $q$.

 Considering the case where the centre of $\mathcal{C}$ is above $q$, it is sufficient to observe that, as $q$ is the closest point on the hyperbola to $s$, then the segment $sq$ is perpendicular to the tangent of the hyperbola at $q$. The gradient of the tangent is  negative, and so as the centre of $\mathcal{C}$ moves upwards from $q$, the distance from the centre to the nearest point on the boundary becomes increasingly greater than $d(q,s)$. 

Now consider $\mathcal{C}$ as its centre adopts some position $q'$ on the hyperbola between $q$ and the line segment $p_0p_1$, as depicted in Figure \ref{fig_gradient_proof_b}. Define $s'$ as the point on $\mathcal{C}'$ intersecting the line $p_0q'$. As long as the gradient of the hyperbola at any point between $q$ and $q'$ is less than the gradient of the arc at any point between $s$ and $s'$, then it is impossible for any intersection to occur between  $\mathcal{C}$ and $\mathcal{C}'$.

The point $x$ is marked as the point on the central axis closest to the point $q'$, i.e. where the points $p_0, q', x$ and $p_1$ are co-circular. The point $s''$ is marked as the point on $\mathcal{C}'$ such that the lines segments $q'x$ and $p_0s''$ are parallel.
The gradient of the tangent of the hyperbola at $q'$ is thus equal to the gradient of the tangent of $\mathcal{C}'$ at $s''$.

It is now sufficient to note that the gradient of $q'x$ is less than that of $p_0x$,  due to the co-circularity of $p_0,q',x$ and $p_1$. Therefore $s''$ is below $s'$ on $\mathcal{C}'$, in all cases where $q$ and $x$ are above the line segment $p_0p_1$. Therefore $\mathcal{C}$ and $\mathcal{C}'$ can never intersect, for any value of $\tau$.

\end{document}